\documentclass[11pt]{article}
\usepackage{amsmath,epsfig,url}
\usepackage{subcaption}
\begin{document}
\baselineskip=16.5pt

\newcommand{\sn}{\sum_{i=1}^n}
\newcommand{\snj}{\sum_{j=1}^n}
\newcommand{\intt}{\int_0^t}
\newcommand{\inti}{\int_0^{\infty}}
\newcommand{\inttau}{\int_0^{\tau}}

\begin{center}
{\Large \bf  Predictive directions for individualized treatment selection in clinical trials} \\
\end{center}

\vspace{.001mm}
\begin{center}
Debashis Ghosh$^1$ and Youngjoo Cho$^2$\\
$^1$Department of Biostatistics and Informatics\\
Colorado School of Public Health, Aurora, CO, 80045\\
debashis.ghosh@ucdenver.edu\\
$^2$Zilber School of Public Health\\
University of Wisconsin-Milwaukee, Milwaukee, WI 53201 USA\\
cho23@uwm.edu
\end{center}

\vspace{.001mm}

\begin{center}
{\large \bf \ Summary}
\end{center}

\vspace{5mm}

\noindent  In many clinical trials, individuals in different subgroups have experience differential treatment effects.  This leads to individualized differences in treatment benefit.  In this article, we introduce the general concept of predictive directions, which are risk scores motivated by potential outcomes considerations.    These techniques borrow heavily from sufficient dimension reduction (SDR) and causal inference methodology.   Under some 
conditions, one can use existing methods from the SDR literature to estimate the directions assuming an idealized complete data structure, which subsequently yields an obvious extension to clinical trial datasets.  In addition, we generalize the direction idea to a nonlinear setting that exploits support vector machines.   
The methodology is illustrated with application to a series of colorectal cancer clinical trials.   
\vspace{5mm}

\noindent {\it Keywords:} Causal effect; heterogeneity of treatment effect; machine learning; model misspecification; personalized medicine; single-index model.
\newpage

\section{{ Introduction}}

\noindent  In many clinical trial settings, the overall treatment effect is the estimand of primary scientific interest, but it may not be appropriate for all the populations considered in the study.  A practical example that is routinely used in clinical practice is testing for DNA variation in the CYP2C19 gene.  For certain variants (e.g., CYP2C19*2, *3, and *17), it has been shown that variation in these single-nucleotide polymorphisms  can be informative about patients' ability to metabolize CYP2C19 substrate drugs \cite{lee2012}.   

With this pharmacogenetic example in mind, developing methods for identification of appropriate patient subgroups for which the treatment might be of major benefit has become a topic of intense interest in the statistical literature.  Gail and Simon \cite{gs1985} introduced methods for identification of qualitative treatment covariate interactions. The Subpopulation Treatment Effect Pattern Plot (STEPP) was developed by Bonetti and Gelber \cite{stepp} as a graphical summary for subgroup identification with attendant permutation testing procedures.   Using a working model and training/test set paradigm, Cai \cite{cai2011} developed a modelling strategy to identify subgroups of patients who would benefit from the treatment; we comment on their approach in \S 3.2.  Tree-based and related machine learning approaches (e.g., \cite{kehlulm,su2008,su2009,foster2011,ir2013,wa2017}) for finding treatment subgroups have also been proposed.  

Much of these methodologies have been focused on the issue of identification of subgroups at a subpopulation level, where the subgroups are defined based on covariates that have interactions with treatment.   Vanderweele and coauthors \cite{vdw2018} took this notion to a person-specific level and described four problems in personalized medicine.  They showed that for each question, the optimal rule has a form that takes the difference in individual-specific responses conditional on covariates.   They use the potential outcomes framework \cite{rubin1974,holland1986} to derive these results.  An important takeaway from their work is the necessity of moving away from testing individual treatment-covariate interactions towards wholistic testing of multiple interactions simultaneously.  

In this work, inspired by ideas from causal inference and its links with sufficient dimension reduction (SDR) methods \cite{ghosh2011,lzg2017}, we develop a concept termed the predictive direction.   The idea is to posit potential outcomes for the subject under each of the possible treatments and to then model their difference.  In the hypothetical case where the complete potential outcomes are available, we can then exploit sufficient dimension reduction methods in order to estimate the predictive direction.  One of the appealing features of such procedures is that in the linear case the estimated predictive direction has an intuitive interpretation as a risk score, i.e., a linear combination of the predictor variables.   Risk scores are commonly used and applied throughout medicine \cite{tripod}.    

While we describe the predictive directions concept within the potential outcomes framework in Section 2, for most situations, the counterfactuals are never simultaneously observed.  To deal with this, we impute the outcomes using random forests \cite{rf2001}, a step that was also applied in the `virtual twins' method of Foster et al. \cite{foster2011}.    In the linear case, a sufficient condition that is needed amounts to effectively a continuous multivariate normality-type assumption on the covariates, which is unlikely to hold in practice.  Thus, we also propose a nonlinear version of the predictive direction whose estimation relies on the use of support vector machines.   The structure of the paper is as follows.  In Section 2, we outline the background material on the potential outcomes framework as well as computation of the predictive direction using SDR methodology.    Section 3 describes a new nonlinear extension of the approach to relax the linearity assumption and yields approximation using kernel machine methods \cite{liu2007}.     Section 5 features an illustration of the techniques to data from 12 colorectal cancer studies we have previously analyzed \cite{ghosh2012}.  Some discussion concludes Section 6.   

\section{{Proposed Framework}}
\subsection{Potential outcomes framework and applications to risk modelling}

We work within the potential outcomes framework of Rubin \cite{rubin1974} and Holland \cite{holland1986}.  
Assume that $(Y_i(0),Y_i(1),T_i,{\bf Z}_i)$, $i=1,\ldots,n$, a
random sample from the triple $(Y(0),Y(1),T,{\bf Z})$, where $(Y(0),Y(1))$ 
represents the counterfactuals, $T$ denotes the treatment group,
and $Z$ is a $p$-dimensional vector of covariates, is observed for all
subjects.  Let 
$T$ take the values $\{0,1\}$ so that
the treatment is binary.  Note that we are merely using the setup
to be able to define the predictive directions.     Also, we will be working within the context
of a clinical trial where $T$ will be randomized so that it can be assumed
to be independent of ${\bf Z}$.  

As described in Rosenbaum and Rubin \cite{rr1983}, the standard assumption needed for causal inference is
that 
\begin{equation}\label{unc}
T \perp \{Y(0),Y(1) \} | {\bf Z},
\end{equation}
i.e. treatment assignment is conditionally independent of the set
of potential outcomes given covariates.  Rosenbaum and Rubin \cite{rr1983} 
refer to (\ref{unc}) as the strongly ignorable treatment assumption; it allows
for the estimation of causal effects.  

In Rosenbaum and Rubin
\cite{rr1983}, the propensity score was introduced as a central quantity needed for the
estimation of causal effects in observational studies.  The propensity score,
defined as the probability of receiving treatment as a function of covariates, is
given by 
\begin{equation}\label{ps}
e({\bf Z}) = P(T = 1|{\bf Z}).
\end{equation}
Rosenbaum and Rubin \cite{rr1983} showed that use of the propensity score leads to
theoretical balance in covariates between the $T = 0$ and $T = 1$ groups.
Statistically, this corresponds to the conditional independence of
$T$ and ${\bf Z}$ conditional on $e({\bf Z})$ and is summarized in Theorem
1 of Rosenbaum and Rubin \cite{rr1983}.  Given the treatment ignorability
assumption in (\ref{unc}), it also follows by Theorem 3 of Rosenbaum and Rubin \cite{rr1983} that treatment is
strongly ignorable given the propensity score, i.e.
$$ {\bf Z} \perp
\{Y(0),Y(1)\} | e({\bf Z}).$$  

We now exploit the work of Ghosh \cite{ghosh2011} and use further conditional independence assumptions from the sufficient dimension reduction literature.   Assume that
there exists a $p \times q$ matrix {\bf A}, $q \leq p$, such that treatment is conditionally independent of 
{\bf Z}, given ${\bf A}'{\bf Z}$.  This can be expressed notationally as 
\begin{equation}\label{ci1}
T \perp {\bf Z }| {\bf A}'{\bf Z}
\end{equation}
Assumption (\ref{ci1}) is a crucial one for defining the estimand targeted by most dimension
reduction methods.  In particular, if $S({\bf A})$ represents the subspace generated by the
columns of ${\bf A}$, then the smallest subspace containing all possible spaces is known as the
central subspace \cite{cook1998} and typically exists in most problems.  

Combining assumptions (\ref{ci1}) and (\ref{unc}), we have  
\begin{equation}\label{unc2}
T \perp \{Y(0),Y(1) \} | {\bf A}'{\bf Z}
\end{equation}
so that the columns of {\bf A} capture the essential information about the potential outcomes.  These columns are what
we term the {\it directions} in the outcome data.   Note that (\ref{unc2}) implies that
\begin{equation}\label{unc3}
T \perp g(\{Y(0),Y(1) \}) | {\bf A}'{\bf Z}
\end{equation}
for any function $g(y,z)$ whose domain is $R^2$ and whose range is $R$.  We now define the function: 
$$ g(y,z) = y-z.$$
Of course, many other functions are possible, but in the current article, we focus on this choice of $g$.    We then define the columns of ${\bf A}$ corresponding to $g$ as the predictive directions.

\subsection{Computation of Predictive Directions}

As noted by Ghosh \cite{ghosh2011}, with the sequence of conditional assumptions being invoked in \S 2.1., one can then employ sufficient dimension reduction procedures in order to compute the predictive directions.   The following high-level algorithm uses sliced inverse regression \cite{li1991}, although other methods could also be used, such as SAVE \cite{cw1991} and MAVE \cite{xia2002}:
\begin{itemize}
\item[A.] Compute $Y^*_i \equiv g\{Y_i(1),Y_i(0)\}$ for subject $i$, $i=1,\ldots,n$.  
\item[B.] Perform sliced inverse regression of $Y^*_i$ on ${\bf Z}_i$ $(i=1,\ldots,n)$ in order to estimate the directions (i.e., the columns of {\bf A}). \end{itemize}
We recall that SIR requires the linearity condition for its validity.  This assumption can be mathematically expressed as $ E({\bf Z}|{\bf A}'{\bf Z}) = {\bf A}'{\bf Z}.$  The linearity condition is viewed as restrictive, as it is effectively satisfied by elliptically symmetric distributions.   

As pointed out before, in practice, we cannot implement the high-level algorithm in the previous paragraph due to the inability to observe both potential outcomes.  Instead of $\{Y_i(0),Y_i(1)\}$, we observe $Y_i = T_iY_i(1) + (1-T_i)Y_i(0)$.   We thus modify the algorithm by including an imputation step: 
\begin{enumerate}
\item Fit a random forests model \cite{rf2001} for $Y_i$ as a function of $T_i,{\bf Z}_i, T_i{\bf Z}_i$, $i=1,\ldots,n$.  
Such an algorithm will allow for computation of $\{\hat Y_i(1),\hat Y_i(0)\}$ based on the observed covariates ${\bf Z}_i$, $i=1,\ldots,n$.
\item Compute the variable $\tilde Y_i = g\{\hat Y_i(1),\hat Y_i(0)\}$ for subject $i$, $i=1,\ldots,n$.
\item Sort $\tilde Y_1,\ldots,\tilde Y_n$ into increasing order and group them into $d$ slices, termed $S_1,\ldots,S_d$.   
\item Standardize the predictor observations as $$\tilde {\bf Z}_i =
\hat \Sigma^{-1/2}({\bf Z}_i - \hat \mu), (i=1,\ldots,n),$$ where $\hat
\mu$ and $\hat \Sigma$ are the sample mean and covariance matrices
of $Z_1,\ldots,Z_n$.
\item Calculate within-slice estimates of sample mean 
$ \bar {\bf Z}_j = {n_j}^{-1} \sum_{i=1}^n I(\tilde Y_i \in S_j)\tilde {\bf Z}_i,$
where $n_j = \sum_{i=1}^n I(\tilde Y_i \in S_j)$, $j=1,\ldots,d$.  
\item Estimate the population covariance matrix of ${\bf Z}$ as 
$$ \hat \Theta = \sum_{j=1}^d \frac{n_j}{n} \bar {\bf Z}_d \bar {\bf Z}'_d.$$
\item Calculate the eigenvalues of $\hat \Theta$.  The estimated directions are the corresponding eigenvectors.  
\end{enumerate}
We make several remarks about this algorithm.  First, since the data come from a randomized clinical trial, separate prediction within treatment arms is a valid approach for imputing potential outcomes.   Second, the approach is agnostic to the choice of imputation algorithm in the first step; one could use other alternatives (e.g., \cite{raghu2001,vb2012}).   Third, step 1 corresponds to the imputation step that is needed in algorithms such as the `virtual twins' algorithm of \cite{foster2011};  however, their subsequent steps are different from ours.    Fourth, one appealing feature of the algorithm is that at the end, we are able to construct risk scores whose coefficients are the eigenvectors, so they enjoy an appealing interpretation from a clinical point of view.   Fifth, an implicit parameter in the algorithm is the number of slices we need to use in step 2.   As Li \cite{li1991} argues, SIR is relatively insensitive to the number of slices used in the algorithm.  Finally, in the current manuscript, we simply use the first eigenvector as our summary predictive direction measure.  Equivalently, we are treating the dimension of the central subspace as being one.  While there is a literature on methods for estimating dimension of the central subspace (e.g., \cite{yeweiss2003,hc2017}), how to use multiple risk scores as well as estimating subspace dimension for the purposes of treatment selection remains an open topic and one that we leave to future investigation.   

\section{Computation of nonlinear predictive directions}

\subsection{A link between SDR with kernel machines}
A crucial assumption in the previous section for the validity of the SDR methodology is the linearity assumption.  In practice, this typically means that the unconditional distribution of ${\bf Z}$ has to have a multivariate normality or related distribution.   There has been much work on developing alternative estimation procedures that seek to relax the linearity assumption.   For example, Xia et al. \cite{xia2002} propose the minimum average variance estimation procedure, which relies on a combination of nonparametric smoothing with weighted least squares.   Since it involves nonparametric regression, its convergence depends on an appropriate rate of convergence for the bandwidth in conjunction with the sample size converging to infinity.   Cook and Ni \cite{cn2005} proposed a minimum discrepancy method in which sufficient dimension reduction is characterized using an objective function approach.   This leads to an alternating least squares algorithm for estimation of the central subspace.  

A seemingly different regression model that could be fit to these data is 
\begin{equation}\label{model1}
Y_i=\beta_0 + h({\bf Z}_i)+\epsilon_i,
\end{equation}
where $\beta_0$ is an intercept term, $h({\bf Z}_i)$ is an unknown centered smooth function,
and the error term $\epsilon_i$ $(i=1,\ldots,n)$ is assumed to be a random sample from a 
$N(0,\sigma^2)$ distribution.   

The kernel machine methodology assumes that $h(\cdot)$ lies in a reproducing Kernel 
Hilbert space \cite{ar1950,wahba1990,bta2004}.  This is a Hilbertian function space ${\cal H}_K$ that satisfies
the property that for any function in ${\cal H}_K$, its pointwise evaluation is a continuous linear functional.
As shown in \cite{ar1950}, there exists a one-to-one correspondence between ${\cal H}_K$ with a so-called kernel function $K({\bf z},{\bf z}^*)$ is a bounded, symmetric, positive function
satisfying
\begin{equation}
\int K({\bf z},{\bf z}^*)h({\bf z})h({\bf z}^*)d{\bf z} d{\bf z}^*\geq 0,
\label{eq:kernel}
\end{equation}
for any arbitrary square integrable function $h({\bf z})$
and all ${\bf z}, {\bf z}^*
\in R^p$.  The kernel function can be viewed as a measure of
similarity between two values of the covariate vector ${\bf z}$ and
${\bf z}^*$. 

Any function $h({\bf z})$ in the function space ${\cal H}_K$
defined by a kernel $K(\cdot,\cdot)$ can have a primal
representation directly using the  basis functions (features) of
${\cal H}_K$, and it can equivalently have a dual representation
using the kernel function $K({\bf z},{\bf z}^*)$ directly.
Specifically, for an arbitrary function  $h({\bf z})\in {\cal
H}_K$, its primal representation takes the form
\begin{equation}
h({\bf z})=\sum_{j=1}^{J}\omega_j\phi_j({\bf z})=\phi({\bf z})^T\omega,
\label{eq:primalrep}
\end{equation}
where $\phi(\cdot)=\{\phi_1(\cdot),\cdots, \phi_J(\cdot)\}^T$
is a $J\times 1$ vector of the standardized orthogonal basis
functions (features), i.e., standardized Mercer features of the
function space ${\cal H}_k$, and the
$\omega=(\omega_1,\cdots,\omega_J)'$ is a vector of some
constants. The square norm of $h(\cdot)$ can be written as
\begin{equation}
\| h \|_{{\cal H}_K}^2
=\sum_{j=1}^{J}\omega_j^2=\omega^T\omega.
\label{eq:normrep}
\end{equation}
Alternatively, the same $h({\bf z})$ can be equivalently written
in a dual representation using the kernel function $K(\cdot,\cdot)$ directly as
\begin{equation}
h({\bf z})=\sum_{l=1}^{L}\alpha_l K({\bf z}_l^*,{\bf z}),
\label{eq:dualrep}
\end{equation}
for some integer $L$, some constants $\alpha_l$ and
some $\{{\bf z}_1^*,\cdots,{\bf z}_L^*\} \in R^p$.
Justifications of these results and more details about the
RKHS can be found in Chapter 3 of \cite{svmbook}.

Exploiting a primal/dual equivalence from Karush-Kuhn-Tucker theory, one can show that the estimator of the
nonparametric function $h(\cdot)$ evaluated at the design points
$({\bf Z}_1, \cdots,{\bf Z}_n)^T$ is estimated as
\begin{equation}\label{hest}
\widehat{{\bf h}}=\lambda^{-1}K(I+
\lambda^{-1}K)^{-1}{\bf y},
\end{equation}
where ${\bf y} \equiv (y_1,\ldots,y_n)$.  
In \cite{liu2007}, it was shown that the estimates of $h$ in (\ref{hest}) can be derived
as arising from a random effects model of the following form: 
\begin{equation}
\label{mix}
{\bf y}= {\bf h}+{\bf e},
\end{equation}
where ${\bf h}$ is an $n\times 1$ vector of random effects following ${\bf h}\sim
N({0},\tau K)$, $\tau$ is a scale parameter,
 and $e\sim
N({\bf 0},\sigma^2{\bf I})$. 
Because of this equivalence, all regression parameters in the model can be estimated by maximum likelihood,
while the variance component parameters can be estimated by restricted maximum likelihood.

Our approach is to link sufficient dimension reduction approaches with the kernel machine methodology that was developed in \cite{liu2007}.  This is done using results from Schoenberg \cite{schoen1938}.   As in Schoenberg \cite{schoen1938}, we will study spaces of positive definite functions that are defined on proper metric spaces.   The space $R^p$ with the Euclidean distance can also be viewed as a metric space.  Let ${\cal B}(E)$ denote the space of positive definite functions for a metric space $E$.   One result of Schoenberg \cite{schoen1938} was that if $E_1$ and $E_2$ are metric spaces with $E_1 \subset E_2$, then ${\cal B}(E_1) \supset {\cal B}(E_2)$.   If we take $E_1$ to be the restriction of $R^p$ to random vectors ${Z}$ that satisfy the linearity condition and $E_2$ to be random vectors which are elliptically symmetric, then we have ${\cal B}(E_1) \supset {\cal B}(E_2)$.  For ${\cal B}(E_2)$, we have the following characterization from Schoenberg \cite{schoen1938}: 

\noindent {\bf Lemma 1.}   A $p-$dimensional random vector ${\bf W}$ is elliptically symmetric if and only if its characteristic function can be written as $\psi(\|{\bf w}\|^2)$, where ${w} \in R^p$ and $\psi(t)$ has the form
\begin{equation}\label{es}
 \psi(t) = \int_0^{\infty} \omega_p(r^2t) dF(r),
 \end{equation}
where $\omega_p$ is the characteristic function for a $p-$dimensional random vector that is distributed uniformly on the unit sphere in $R^p$, and $F(r)$ is a distribution function on $[0,\infty)$.  We note that the form of $\omega_p(t)$ is given by
$$ \omega_p(t) = \Gamma \left ( \frac{p}{2} \right ) \left ( \frac{2}{t} \right )^{(p-2)/2} J_{(p-2)/2}(t),$$
where $\Gamma(a) \equiv \int_0^{\infty} u^{a-1}\exp(-u) du $ denotes the Gamma function and 
$$J_{\alpha}(x) \equiv \sum_{m=0}^{\infty} \frac{(-1)^m}{m!\Gamma(m+\alpha+1)} \left ( \frac{x}{2} \right )^{2m+\alpha}$$
represents the Bessel function.   

Given the definitions of $E_1$ and $E_2$, we define a sequence of metric spaces in the following way: define $E_{2+i}$ is a metric space consisting of elliptically symmetric random vectors in $R^{p+i}$ for $i=1,2,\ldots,$.   We have that elliptical symmetry in higher dimensions imply elliptical symmetry in lower dimensions.  This implies the following chain of inequalities:
\begin{equation}\label{inequalities}
{\cal B}(E_1) \supset {\cal B}(E_2) \supset {\cal B}(E_3) \supset \cdots \supset  {\cal B}(E_{\infty}).
\end{equation}
In addition, Schoenberg \cite{schoen1938} provides a characterization of ${\cal B}(E_{\infty})$ in (\ref{inequalities}), which is given in the following result:

\noindent {\bf Lemma 2.} A random element ${\cal W}$ exists in ${\cal B}(E_{\infty})$  if and only if its characteristic function can be written as 
$\psi(\|{w}\|^2)$, where $\psi(t)$ has the form
\begin{equation}\label{eshilbert}
 \psi(t) = \int_0^{\infty} \exp(-r^2t) dF(r), \ \ t >0
 \end{equation}
and $F(r)$ is a distribution function on $[0,\infty)$.  

\noindent {\bf Remark 1.}   Note that by the nested structure of the space of positive definite functions in (\ref{inequalities}), it is also the case that
$${\cal B}(E_{\infty}) = \cap_{i=1}^{\infty} {\cal B}(E_i).$$
Thus, ${\cal B}(E_{\infty})$ is the smallest space containing ${\cal B}(E_i)$ for all $i$.  In this sense, the object ${\cal B}(E_{\infty})$ can be interpreted as an infinite-dimensional analog to the central subspace that was described in \S 2.1.   A different type of limiting object corresponding to the central subspace in a nonlinear setting has been developed by Lee et al. \cite{lee2013}.

For our proposed methodology, we will require the definitions of positive definite and completely monotone functions.   
\\
\noindent {\bf Definition 1.}  A real-valued function $f$ is said to be positive definite if for any set of real numbers $x_1,\ldots,x_n$, the $n \times n$ matrix {\bf A} with $(i,j)$th entry $a_{ij} = f(x_i - x_j)$ $(i=1,\ldots,n; j=1,\ldots,n)$ is positive definite.   
\\
\noindent {\bf Definition 2.} A real-valued function $f$ is said to be completely monotone if for all $r \in \{0,1,2,\ldots\}$, 
$$ (-1)^r f^{(r)}(x) \geq 0,$$
where $f^{(r)}$ denotes the $r-$the derivative of $f$.   

A function $f(t)$ $(t \in R)$ is positive definite if and only if $f(t) = g(t^2)$, where $g$ is completely monotone.  The other key fact is that any positive definite function can define a kernel .  Thus, for any positive definite function  $f$, we have that $K({\bf Z},{\bf Z}^*) = f(\|{\bf Z} - {\bf Z}^*\|)$ is a proper kernel.   
Combining these results, we have the following.   

\noindent {\bf Proposition.}   A random element exists in ${\cal B}(E_{\infty})$ if and only if its associated kernel is of the form 
\begin{equation}\label{tinv}
K({\bf X},{\bf X}^*) = \psi(\|{\bf X} - {\bf X}^*\|),
\end{equation}
where $\psi$ is generated via (\ref{eshilbert}).   

The proposition shows that the kernels in ${\cal B}(E_{\infty})$ only depend on the interpoint distances between points.

\noindent {\bf Remark 2.}  Each element of ${\cal B}(E_{\infty})$ will have a unique kernel associated with it and vice versa.   One example of a kernel that would exist in ${\cal B}(E_{\infty})$ is the Gaussian Kernel, whose kernel is given by 
$$K({\bf z},{\bf z}^*)= {\rm
exp}\{-\|{\bf z}-{\bf z}^*\ \| ^2/\rho\},$$
 where
$\|{\bf z}- {\bf z}^*\| =\{\sum_{k=1}^p (z_k-z^*_k)^2\}^{1/2}$. The Gaussian
kernel generates the function space spanned by radial basis
functions, a complete overview for which can be found in \cite{rbfbook}.   Other examples of kernels that reside in ${\cal B}(E_{\infty})$ can be found in Table 1. 

\begin{center}
[\bf Table 1. about here.]
\end{center}

\subsection{Proposed Algorithm and tuning parameter selection}

The results in the previous section lead to a modification of the algorithm in \S 2.2.   It now proceeds as follows:
\begin{enumerate}
\item Fit random forests for $Y_i$ as a function of $T_i,{\bf Z}_i,$ and $T_i{\bf Z}_i$, $i=1,\ldots,n$.  
Such an algorithm will allow for computation of $(\hat Y_i(1),\hat Y_i(0))$ based on the observed covariates ${\bf Z}_i$, $i=1,\ldots,n$.
\item Compute the variable $\tilde Y_i = g\{\hat Y_i(1),\hat Y_i(0)\}$ for subject $i$, $i=1,\ldots,n$.
 \item Fit a Gaussian kernel machine model to $\tilde Y_i$ as a function of ${\bf Z}_i$, $i=1,\ldots,n$.
\end{enumerate}
One then gets fitted values from the kernel machine applied to the input covariate vectors, and these can be treated as functionals of nonlinear extensions of the predictive directions defined in \S 2.1.    Note that the third step amounts to fitting a support vector regression models, details of which can be found in Chapter 6 of Cristianini and Shawe-Taylor \cite{svmbook}.  We use the {\bf svm} function in the {\bf e1071} package in the R library to fit this.  

A natural question that arises is how to set tuning parameters in the Gaussian kernel machine in Step 3.   We follow the advice of Athey and Imbens \cite{ai2016} and divide the training dataset in Step 3 into two independent parts.  For the first part, we optimize the kernel machine to find optimal tuning parameters; this is done using cross-validation.  Given the optimal tuning parameters, we then fit the kernel machine in step 3 with the optimized parameters.  

For many situations, we might wish to perform evaluations on a test set, as discussed in the next section.    For that case, we argue that the tuning parameter selection is less of an issue.    In the parlance of Cai et al. \cite{cai2011}, this is a working model that is used in the direction estimation algorithm.    If our final evaluations are performed on an independent test set, we can argue as in \cite{cai2011} that the ultimate estimands of interest do not rely on proper specification of the working model and therefore enjoy a certain robustness property.  

We note that a related approach to using kernel machines was taken in Shen and Cai \cite{shencai2016}.  While their approach shares similarities with the algorithm developed here, we note that the motivation and starting points are completely different.   Furthermore, they were focused more on the issue of testing, while our goal here is that of computing and estimating directions.

\section{Optimality of treatment selection rules and evaluation of predictive directions}

Based on our approaches to predictive direction estimations, we can now use the directions to guide optimal treatment strategies using the framework developed in \cite{vdw2018}.  For these four questions, that the optimal rule is to treat those subjects for whom $Y(1) - Y(0) > k$, where $k$ get chosen in a context-dependent way.  Since $Y(1) - Y(0)$ does not get observed, our proxy rule is to instead use
\begin{equation}\label{optrule}
D_{10} > k,
\end{equation}
where $D_{10}$ is the predictive direction-derived score.    

To evaluate the predictive direction as a scoring rule, we need a training and testing set in which both studies are randomized and consist of the same treatments.  In addition, outcome variables need to be measured in both studies.   The proposal is related to one discussed in Vickers et al. \cite{vickers2007}.  To simplify the discussion, we will deal with the case of two treatment groups.  The procedure works as follows:
\begin{itemize}
\item[(a).] Estimate the predictive direction using the training dataset.
\item[(b).] Using the estimated direction, compute scores for all subjects in the test set.
\item[(c).] Based on the scores, determine which treatment each subject should receive in the test set using treatment rules of the form (\ref{optrule}).   
\item[(d).] For the subjects whose predicted treatment match their randomized treatment in the test set, compare the outcomes between the two treatment groups. 
\end{itemize}
We mention some points at this stage.  First, we note that for step (b), the outcome information in the test set is not used at all.  Only the covariate information is used to compute the scores.   The outcome information is needed in step (d). in order to compute the measure of treatment effect between the two groups.   Note also that the fact that the test set also comes from a clinical trial is a necessary feature here.    In step (d)., we will be excluding two types of subjects in the test set: those who were predicted to have greatest benefit from one treatment group but were observed to receive the other one.    Thus, we are performing a subgroup analysis in step (d). based on subjects in the test set whose predicted and actual treatment assignments are concordant.  The randomization of treatment is necessary in order to ensure that the subgroup analysis will also be the same as the overall treatment effect.   

\section{{Meta-analysis of colorectal cancer datasets}}

In this section, we will apply the proposed methods to data from a series of 12 adjuvant colon cancer studies that were evaluated for surrogacy in Ghosh et al. \cite{ghosh2012}. Here, we will use data on treatment, age at baseline, stage and gender to explore predictive directions with respect to survival time.   The original 12 studies sought to evaluate the difference in survival times between treatments.  Note that in our previous discussion, we assumed that the variable of interest is continuous.   In the colorectal cancer dataset, the endpoint of interest is time to death.    With respect to the Vanderweele et al. \cite{vdw2018} framework, we are dealing with their second question: given no resource constraints, who should we treat?

We perform a simple modification of the algorithms presented by following a suggestion from Keles and Segal \cite{ks2002}.  We compute a first-stage martingale residual from a null model (i.e., one with no covariates).  We then treat the residual as a continuous variable to be input into the algorithms in \S 2.2. and \S 3.    In addition, because we have data on 12 studies, we can furthermore explore the issue of whether or not the estimated directions show concordance across studies.   

Using SIR, we compute the predictive directions and assess their concordance across the 12 studies.   The results are shown in Figure 1.  

\begin{center}
[\bf Figure 1. about here.]
\end{center}
Based on the plot, we find that there is a relative lack of concordance in terms of the effects of the covariates across the different studies. This suggests the difficulty in finding such interactions as well as in the lack of replicability of interactions across studies.

Next, we evaluated the fitted values using the procedure from \S 4.  Each study was used as a training dataset, with the remaining 11 studies used as test dataset.   The results from using the sliced inverse regression-based procedure is shown in Table 2.

\begin{center}
[\bf Table 2. about here.]
\end{center}

While the studies suggest that the linear predictive directions lead to some benefit of selecting patients in a consistent, we mention two things at this point.   First, for half of the studies, we were unable to compute a hazard ratio.  This was due to the fact that the estimated predictive directions did not lead to predicted treatment assignments that were concordant with the observed treatment assignments in the test datasets.   These analyses suggest that the predictive direction is not generalizable from the series of 12 colorectal cancer trials.  

We now redo the analyses using the nonlinear methodology from \S 3.  Based on the support vector machine, the estimated directions are given in Figure 2.  
\begin{center}
[\bf Figure 2. about here.]
\end{center}
Much like Table 2, this figure shows the high degree of variation across study.  This again speaks to the capacity of being able to find a generalizable person-specific interaction effect for this setting.  

Finally, we ran the prediction analysis similar to what was described in Table 2.  We performed both with and without split-sample optimization.   The results are given in Table 3.
\begin{center}
[\bf Table 3. about here.]
\end{center}
One thing to note is that we are now able to estimate hazard ratios and confidence intervals for all studies and no longer suffer from the numerical issues in the linear case.  However, we again see heterogeneity in the treatment effect  across studies.    In addition, there appears to be little difference between the estimated effects and inference based on whether perform the split-sample optimization or not.   Two exceptions appear to be studies C04 and C07, where the direction of the effect reverses based on whether or not split-sample optimization is performed.   However, both studies are also consistent with no difference between the treatment groups.   

\section{Discussion}
In this article, we have developed the concept of predictive directions for identification of person-specific effects in clinical trials.   In the linear case, we are able to obtain linear combinations of the covariates that enjoy a risk score interpretation.  However, the validity of predictive directions requires strong distributional assumptions, so we have also proposed a novel nonlinear extension that applies support vector regression techniques.   Based on the numerical issues seen in the example in \S 5, we would argue for use of the nonlinear approach, which has also been advocated by other proponents in the SDR literature (e.g., \cite{lee2013,li2011}).  

There are several potential extensions of this work that are currently under investigation.   First, the issue of dimension estimation and subsequent post-model selection inference has not been addressed.  In the current paper, we have bypassed the issue by fixing the dimension to be one.   In the situation where there are multiple directions (i.e., multiple columns of {\bf A} in (\ref{unc3})), a natural question arises as to how to use them to inform selection of optimal treatment as discussed in \S 4.   Finally, a more direct extension to the survival data example in \S 5 would have used the random forests methodology for survival data \cite{rsf2008}.   However, that use of that framework would then require rephrasing the potential outcomes model and attendant assumptions in \S 2.1.   

\section*{Acknowledgments}
This research is supported by National Institutes of Health grant CA129102.

\newpage

\section*{Figures and Tables}

\begin{table}[htbp!]
\begin{center}
\caption{Examples of kernels that are members of ${\cal B}(E_{\infty})$.  Here $K_{\nu}$ denotes the modified Bessel
function of the second kind of order $\nu$.  }
\begin{tabular}{l l l}
\\
\hline
Kernel	&  K$({\bf z},{\bf z}^*)$	& Parameter ranges \\
\hline
Gaussian& $\exp\{-\|{\bf z}-{\bf z}^*\ \| ^2/\rho\}$ & $\rho > 0$\\
Mat\'ern& $\frac{2^{\nu-1}}{\Gamma(\nu)} \left ({\|{\bf z}-{\bf z}^*\|}/{c} \right)^{\nu} K_{\nu}\left ({\|{\bf z}- {\bf z}^*\|}/{c} \right) $& $c,\nu > 0$  \\
Generalized Cauchy &$\left [ 1 + \left ({\|{\bf z} - {\bf z}^*\|}/{c} \right )^{\alpha} \right ]^{-\tau/\alpha}$ & $c, \tau > 0,  0 < \alpha \leq 2$ \\
Dagum & \\
Powered Exponential & $\exp\{-\left ( {\|{\bf z}- {\bf z}^*\|}/{c}\right)^{\alpha} \}$ & $c > 0, 0 < \alpha \leq 2$  \\
\hline
\end{tabular}
\end{center}
\end{table}

\begin{table}[htbp!]
\centering
\begin{tabular}{rrl}
  \hline
Training Data & HR & 95\% CI  \\ 
  \hline
  C04 & 1.21 & (0.79,1.86) \\ 
  NCCTG-78-48-52 & 0.75 & (0.69,0.82) \\ 
  NCCTG-89-46-51 & 1.15 & (0.85,1.56) \\ 
  NCCTG-91-46-53 & 0.92 & (0.85,1.00) \\ 
  C06 & 0.90 & (0.65,1.23) \\ 
  C07 & 0.66 & (0.60,0.73) \\ 
   \hline
\end{tabular}
\caption{Results from computing hazard ratios for estimated predictive directions.   The column titled `Training Data' denotes the colorectal cancer study that was used to estimate the predictive directions.   The second column denotes the hazard ratio computed using the remaining 11 studies as a test dataset based on the procedure outlined in \S 2.2.  The third column denotes a 95\% confidence interval for the hazard ratio computed using the normal approximation.  }
\end{table}

\begin{figure}[htbp!]
\begin{center}
\epsfig{file=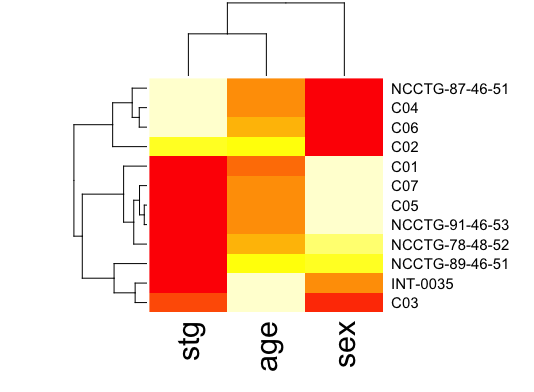,height=2.5in,width=5.5in}
\caption{Heatmap showing the coefficients of the estimated predictive directions across the 12 colorectal cancer studies.  Each row represents a different study, while the columns represent the three variables (sex, age and stage (shown as stg)).   Colors is red represent negative coefficients, while less red denotes more positive coefficients.}
\end{center}
\end{figure}

\begin{figure}[htbp!]
\begin{center}
\epsfig{file=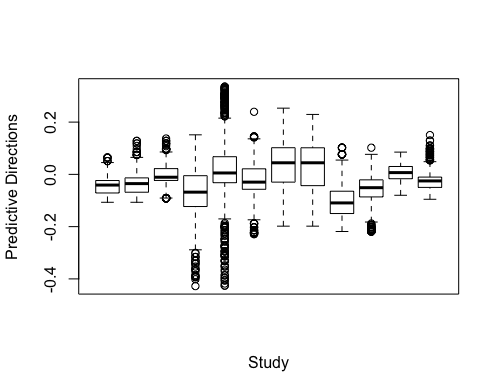,height=2.5in,width=5.5in}
\caption{Boxplot of predictive directions (i.e., the fitted values from the support vector regression approach in \S 3) by study.}
\end{center}
\end{figure}

\begin{table}[htbp!]
\centering
\begin{tabular}{rrlrl}
  \hline
  & \multicolumn{2}{c}{Without optimization} & \multicolumn{2}{c}{With optimization} \\
Training Data & HR & 95\% CI & HR  & 95\% CI  \\ 
  \hline
C01 & 1.01 & (0.93,1.10) & 0.98 & (0.91,1.06) \\ 
  C02 & 1.38 &(1.28,1.49) & 1.26 & (1.16,1.36) \\ 
  C03 & 1.22 & (1.10,1.34) & 1.04 & (0.94,1.15) \\ 
  C04 & 1.15 & (1.03,1.28) & 0.93 & (0.81,1.07) \\ 
  C05 & 1.07 & (0.99,1.16) & 1.00 & (0.92,1.08) \\ 
  INT-0035 & 0.47 & (0.39,0.58) & 0.48 & (0.41,0.56) \\ 
  NCCTG-78-48-52 & 0.72 & (0.66,0.79) & 0.73 & (0.67,0.79) \\ 
  NCCTG-87-46-51 & 1.25 & (1.09,1.45) & 1.38 & (1.20, 1.60) \\ 
  NCCTG-89-46-51 & 0.89 & (0.82,0.96) & 0.84 & (0.78, 0.91) \\ 
  NCCTG-91-46-53 & 0.95 & (0.87,1.03) & 1.00 & (0.91,1.08) \\ 
  C06 & 1.09 & (1.00,1.18) & 1.01 & (0.93,1.09) \\ 
  C07 & 0.90 & (0.80,1.01) & 1.03 & (0.94,1.13) \\ 
   \hline
\end{tabular}
\caption{Results from computing hazard ratios for estimated predictive directions using the methdology
of \S 3.  The column titles are the same as in Table 2.}
\end{table}

\begin{table}[htbp!]
\centering
\begin{tabular}{rrrr}
  \hline
Study & age & stage & sex \\ 
  \hline
C01 & 0.029 & -0.251 & 0.968 \\ 
  C02 & 0.033 & 0.146 & -0.989 \\ 
  C03 & -0.013 & -0.658 & -0.753 \\ 
  C04 & -0.077 & 0.802 & -0.593 \\ 
  C05 & -0.004 & -0.627 & 0.779 \\ 
  INT-0035 & -0.020 & -0.858 & -0.513 \\ 
  NCCTG-78-48-52 & 0.021 & -0.780 & 0.625 \\ 
  NCCTG-87-46-51 & 0.155 & 0.887 & -0.435 \\ 
  NCCTG-89-46-51 & 0.015 & -0.997 & 0.072 \\ 
  NCCTG-91-46-53 & 0.020 & -0.670 & 0.742 \\ 
  C06 & -0.002 & 0.681 & -0.733 \\ 
  C07 & 0.007 & -0.516 & 0.857 \\ 
   \hline
\end{tabular}
\caption{Estimated predictive directions for each of the 12 colorectal cancer studies.  The numbers in each column represent the coefficient for the estimated predictive direction.  For example, the estimated predictive direction for study C01 is given by 
$$ 0.029\text{age}-0.251\text{stage}+0.968\text{sex}.$$}
\end{table}


\begin{thebibliography}{siam}


\bibitem{lee2012} Lee S--J. Clinical Application of CYP2C19 Pharmacogenetics Toward More Personalized Medicine. Frontiers in Genetics. 2012;3:318. doi:10.3389/fgene.2012.00318.

\bibitem{gs1985} Gail M, Simon R. Testing for qualitative interactions between treatment
effects and patient subsets. Biometrics. 1985 Jun;41(2):361-72.



\bibitem{stepp} Bonetti M, Gelber RD. Patterns of treatment effects in subsets of patients in 
clinical trials. Biostatistics. 2004 Jul;5(3):465-81. PubMed PMID: 15208206.

\bibitem{cai2011} Cai T, Tian L, Wong PH, Wei LJ. Analysis of randomized comparative clinical
trial data for personalized treatment selections. Biostatistics. 2011
Apr;12(2):270-82. 


\bibitem{kehlulm} Kehl V, Ulm K.  Responder identification in clinical trials with censored data. {Comput Stat Data Anal}  2006; {50}: 1338 -- 1355.


\bibitem{su2008} Su X, Zhou T, Yan X, Fan J, Yang S. Interaction trees with censored survival data. { Intl J Biostatistics} 2008; {\bf 4}:  article 2.


\bibitem{su2009} Su X, Tsai CL, Wang H, Nickerson DM, Bogong L. Subgroup analysis via recursive partitioning. {J Mach Learn Res} 2009; {10}: 141 -- 158.


 \bibitem{foster2011} Foster JC, Taylor JM,  Ruberg SJ.  Subgroup identification from randomized clinical trial data. {Stat Med}  2011; {30}: 2867 -- 80. 


\bibitem{ir2013} Imai K, Ratkovic M. Estimating treatment effect heterogeneity in randomized program evaluation.  {Ann Appl Stat} 2013; {7}: 443 -- 470.


\bibitem{wa2017} Wager S, Athey S. Estimation and inference of heterogeneous treatment effects using random forests.   {J Am Statist Assoc} 2018; Forthcoming.


\bibitem{vdw2018} VanderWeele TJ, Luedtke AR, van der Laan MJ, Kessler RC.  Selecting optimal subgroups for treatment selection  using many covariates.   Available at 
 \url{https://arxiv.org/abs/1802.09642}.  


\bibitem{rubin1974} Rubin DB. Estimating causal effects of
treatments in randomized and nonrandomized studies.  {J Educ Psych} 1974; {66}: 688 -- 701.


\bibitem{holland1986} Holland P.  Statistics and causal inference (with
discussion).  {J Am Statist Assoc} 1986; {81}: 945 -- 970.


\bibitem{ghosh2011} Ghosh D.  Propensity score modelling in observational studies using dimension reduction
methods. {Stat Prob Lett} 2011; 81:  813  -- 820.


\bibitem{lzg2017}  Luo W, Zhu Y, Ghosh D.  On estimating regression-based causal effects using sufficient dimension reduction. {Biometrika} 2017; {104}:  51 -- 65.  


\bibitem{tripod} Collins GS, Reitsma JB, Altman DG, Moons KG. Transparent reporting of a
multivariable prediction model for individual prognosis or diagnosis (TRIPOD):
the TRIPOD statement. BJOG. 2015;122: 434-43.


\bibitem{rf2001} Breiman L. Random forests.  {Mach Learn} 2001; {45}:  5 -- 32.

\bibitem{liu2007} Liu D, Lin X, Ghosh D. Semiparametric regression of
multi-dimensional genetic pathway data: least squares kernel machines 
and linear mixed models.  {Biometrics} 2007; {63}, 1079 -- 1088.  

\bibitem{ghosh2012} Ghosh D, Taylor JMG, Sargent DJ. Meta-analysis for surrogacy: accelerated failure time modelling and semi-competing risks (with discussion).  {Biometrics} 2012; {68}: 226 -- 247.


\bibitem{rr1983} Rosenbaum PR, Rubin DB.  The central
role of the propensity score in observational studies for causal
effects.  {Biometrika} 1983; {70}: 41 -- 55.



\bibitem{cook1998} Cook RD.  {\it Regression Graphics}.  New York: Wiley, 1998.



\bibitem{li1991} Li KC.  Sliced inverse regression for dimension
reduction (with discussion).  {J Am
Statist Assoc} 1991; {86}: 316 -- 342.


\bibitem{cw1991} Cook RD, Weisberg S. Discussion of `Sliced inverse regression' by Li.  {J Am
Statist Assoc} 1991; {86}: 328 -- 332.

\bibitem{xia2002} Xia Y, Tong H, Li WK, Zhu LX.  An adaptive estimation of dimension reduction space (with discussion).  {J R Statist Soc Ser B} 2002; {64}: 363 -- 410.


\bibitem{raghu2001} Raghunathan TE, Lepkowski JM, Van Hoewyk JV, Solenberger P.  A Multivariate Technique for Multiply Imputing Missing Values Using a Sequence of Regression Models.  {Surv Methodol} 2001; {27}: 85 -- 95.

\bibitem{vb2012}  Van Buuren S. 
\textit{Flexible Imputation of Missing Data}.
Boca Raton, FL: Chapman \& Hall/CRC Press, 2012.

\bibitem{yeweiss2003} Ye Z, Weiss RE.  Using the bootstrap to select one of a new class of dimension reduction methods. { J Amer Statist Assoc} 2003;  {98}: 968 -- 979.

\bibitem{hc2017} Huang MY, Chiang CT.  An effective semiparametric estimation approach for the sufficient dimension reduction model.   {J Amer Statist Assoc}  2017; {112}, 1296 -- 1310. 

\bibitem{cn2005} Cook RD, Ni L.  Sufficient dimension reduction via inverse regression: A
minimum discrepancy approach. {J Amer Statist Assoc} 2005; {100}, 411 -- 428.

\bibitem{ar1950} Aronszajn N. Theory of reproducing kernels. {Trans Am Math Soc} 1950; {68}: 337 -- 404. 

\bibitem{wahba1990} Wahba G.  {\it Spline Models for Observational Data}.  Philadelphia: SIAM, 1990.

\bibitem{bta2004} Berlinet A, Thomas-Agnan C.  {\it Reproducing Kernel Hilbert Spaces in Probability and Statistics.} Kluwer Academic Publishers, 2004. 

\bibitem{svmbook} Cristianini N, Shawe-Taylor J.  {\it An
Introduction to Support Vector Machines and Other Kernel-based
Learning Methods.}  Cambridge: Cambridge University Press, 2000.


\bibitem{schoen1938} Schoenberg IJ.  Metric spaces and completely monotone functions.  {\it Ann  Math} {39}, 811 -- 841.

\bibitem{lee2013} Lee KY, Li B, Chiaromonte F.  A general theory for nonlinear sufficient dimension reduction: Formulation and estimation. {Ann Stat} 2013; {41}: 221--249. 


\bibitem{rbfbook} B\"uhmann MD. {\it Radial Basis Functions: Theory and Implemetations}.   Cambridge: Cambridge University Press, 2003.



\bibitem{ai2016} Athey S, Imbens G. Recursive partitioning for heterogeneous causal effects.  { Proc Nat Acad Sci} 2016; 113: 7353  -- 7360.

\bibitem{shencai2016} Shen Y, Cai T.  Identifying
predictive markers for personalized treatment selection.  {Biometrics} 2016; {72}: 1017 -- 1025. 



\bibitem{vickers2007} Vickers AJ, Kattan MW, Sargent D. Method for evaluating prediction models that 
apply the results of randomized trials to individual patients.  {Trials} 2007; {8}, 14.


\bibitem{ks2002} Keles S, Segal MR.  Residual-based tree-structured survival analysis. {Stat Med} 2002; {21}: 313 -- 326. 

\bibitem{li2011} Li B, Artemiou A, Li L.  Principal support vector machines for linear and nonlinear sufficient dimension reduction. {Ann Stat} 2011; {39}: 3182--3210. 

\bibitem{rsf2008} Ishwaran H, Kogalur UB, Blackstone EH, Lauer MS.  Random survival forests.  {Ann App Stat} 2008; {2}: 841 -- 860.





\end{thebibliography}
\end{document}